\documentclass[12pt]{article} 
\usepackage{epsf}
\usepackage{latexsym}

\usepackage{hyperref} 

\DeclareFixedFont{\xiiss}{OT1}{cmss}{m}{n}{12}
\DeclareFixedFont{\ixss}{OT1}{cmss}{m}{n}{9}
\DeclareFixedFont{\cmrnine}{OT1}{cmr}{m}{n}{9}
 
\newcommand{\CC}{\hbox{\xiiss C\kern-.4emI}}
\newcommand{\RR}{\hbox{\xiiss R\kern-.45emI}}
\newcommand{\ZZ}{\hbox{\xiiss Z\kern-.4emZ}}
\newcommand{\CCs}{\hbox{\ixss C\kern-.4emI}}
\newcommand{\ZZs}{\hbox{\ixss Z\kern-.4emZ}}
\newcommand{\pa}{\partial}
\newcommand{\tpsi}{\tilde\psi}
\newcommand{\bpaX}{\bar\pa X}
\newcommand{\paX}{\pa X}
\newcommand{\tr}{{\rm tr\ }}

\newcommand{\delpp}{\delta^\perp}
\newcommand{\delpa}{\delta^\parallel}

\newcommand{\beq}{\begin{equation}}
\newcommand{\beql}[1]{\begin{equation}\label{eq:#1}}
\newcommand{\eeq}{\end{equation}}
\newcommand{\be}{\begin{equation}}
\newcommand{\ee}{\end{equation}}
\newcommand{\beqn}{\begin{eqnarray}}
\newcommand{\eeqn}{\end{eqnarray}}
\newcommand{\bea}{\begin{eqnarray}}
\newcommand{\eea}{\end{eqnarray}}

\newcommand{\eq}[1]{(\ref{eq:#1})}

\newcommand{\littlefig}[2]{
	\epsfxsize=#2in
	\epsfbox{#1}
}

\begin{document}
        \begin{titlepage}
        \title{
                \begin{flushright}
                \begin{small}
                ILL-(TH)-99-02\\
                hep-th/9904104\\
                \end{small}
                \end{flushright}
			Superstring Perturbation Theory\\ and Ramond-Ramond Backgrounds
		}

\author{ 	David Berenstein\thanks{email:{\tt berenste@hepux0.hep.uiuc.edu}}
			\\
			and \\
			Robert G. Leigh\thanks{email:{\tt rgleigh@uiuc.edu }}\\ 
        	\\
                {\small\it Department of Physics}\\
                {\small\it University of Illinois at Urbana-Champaign}\\
                {\small\it Urbana, IL 61801}\\
		}

        \maketitle

        \begin{abstract}
We consider perturbative Type II superstring theory in the covariant NSR
formalism in the presence of NSNS and RR backgrounds. A concrete example
that we have in mind is the geometry of D3-branes which in the
near-horizon region is $AdS_5\times S_5$, although our methods may be
applied to other backgrounds as well. We show how conformal invariance
of the string path integral is maintained order by order in the number
of holes.
This procedure makes uses of the Fischler-Susskind mechanism to build up
the background geometry. A simple formal expression is given for a
$\sigma$-model Lagrangian. This suggests a perturbative expansion in
$1/g^2N$ and $1/N$. As applications, we consider at leading order the
mixing of RR and NSNS states, and the realization of the spacetime
supersymmetry algebra.
        \end{abstract}

        \end{titlepage}


\section{Introduction}

The quantization of superstrings in non-trivial backgrounds has become
an urgent problem since it has been realized that there is a direct
relationship to the strong coupling dynamics of gauge field theories. In
particular, one is often interested in conformal field theories which
are related to spacetimes which contain an anti-deSitter factor.\cite{Juan}
  For
examples with $AdS_3$ and backgrounds of NSNS fields, several cases have
been worked out\cite{GKS,EFGT,KLL,BORT,newKS} for finite $N$. The more
interesting case with RR backgrounds, corresponding to configurations of
D-branes, are much more difficult. Work on $AdS_3$ has been presented in
Ref. \cite{BVW}, while $AdS_5\times S^5$ has been discussed 
in the Green-Schwarz formalism\cite{Metsaev:1998it,renata}. Ref. 
\cite{ARMR}  discusses the latter in the context of an
expansion around flat spacetime.

In this paper, we consider RR backgrounds in the covariant RNS formalism.
We begin with the theory in flat spacetime and demonstrate, that in the
large $N$ limit, string perturbation theory organizes itself into a
$\sigma$-model which is exactly conformal. This procedure may be thought
of as summing of worldsheet holes, with closed string loops suppressed
because of large $N$. We check conformal invariance
of the string path integral explicitly at leading non-trivial order, an
application of the Fischler-Susskind mechanism.\cite{WFLS}
The $\sigma$-model is of a non-standard type because of the RR background,
but can be written formally in a way which is useful for perturbative 
calculations. 

The case of $AdS_5\times S^5$ is of particular interest. This model has
a great deal of symmetry and it is possible that a free conformal field
theory description exists, perhaps similar to the NS backgrounds with
$AdS_3$. There are a number of oddities, centering around the properties
of superconformal ghosts, as well as the properties of the spacetime
supersymmetry algebra, which remain elusive.\cite{BLL} The present
construction however is certainly not such an exact description, and
must be discussed in the context of a weak curvature expansion. The analysis
may be applied to other cases as well, such as the $D1-D5$ system, but we
confine our attention here to $AdS_5\times S^5$.

The paper is organized as follows. In section 2, we discuss general
properties of string perturbation theory and demonstrate that in the
large $N$ limit, the leading effect corresponds to a sum over worldsheet
boundaries. We argue that these effects can be accounted for by a 
formal $\sigma$-model which includes the RR background. In order
to define the $\sigma$-model it is necessary to introduce an operator
which acts like the square-root of the picture-changing operator, $P_{1/2}$. 
In section 3,
we discuss the conformal invariance of this $\sigma$-model at lowest
order in $\alpha'$ and demonstrate that the Fischler-Susskind mechanism properly
resums the background geometry. In section 4, we discuss some simple
aspects of the spectrum of excitations around this background, and
demonstrate that at a given level, R and NS states mix. In section 5, 
we consider the realization of spacetime symmetry charges in terms of
worldsheet currents and comment on the supersymmetry algebra. Our
analysis here is limited by our incomplete knowledge of the properties
of $P_{1/2}$. Further comments and speculations are reserved for the
final section.

\section{Background of D3-branes: Summing over Boundaries}

The basic idea here for computing the partition function is familiar
from field theory: consider a scalar field theory, with a background
field turned on. The one-loop determinant may be thought of as being 
built up by summing over multiple insertions of the background.
 
\centerline{\littlefig{485oneloopeff.EPSF}{3}}

Indeed, in string theory in NSNS backgrounds, it is well known that
summing over multiple insertions of vertex operators simply
exponentiates those vertex operators, resulting in a $\sigma$-model.
Here, we repeat this analysis in the case of interest, with
RR-backgrounds, and we will discover in exactly what sense the RR vertex
operator exponentiates. We will find that it is possible to write a
formal expression (involving one term) that quantifies the effect of the
RR-background. This is necessarily formal, because it involves an
operator which changes ghost number in a non-standard way. The
alternative is to remove this operator, but at the expense of
introducing an infinite number of non-local operators into the
``$\sigma$-model''. The utility of the formal expression is that it
automatically generates the correct combinatorics, and has a clear
interpretation for any non-zero S-matrix element.

Let us begin by discussing the classical geometry exterior to a
collection of $N$ D3-branes. The background is given by
\beql{bgmetric}
ds^2=f(r)^{1/2}dx_\perp^2+f(r)^{-1/2}dx_\parallel^2
\eeq
where $f(r)=1+R_s^4/r^4$, $r$ being the radial perpendicular coordinate.
The radius of curvature satisfies $R_s^4=4\pi gN\alpha'^2$. In addition,
there is a self-dual RR five-form fieldstrength, with total flux $N$, and
the dilaton is constant. One notes that for large $gN$, the curvature
of spacetime is everywhere small, and therefore solving the equations of
motion at first order in $\alpha'$ is a very good approximation. Furthermore,
it has been argued from spacetime considerations that the near-horizon
geometry is exact to all orders in $\alpha'$ and $g$.

Our first task is to understand these results directly from string 
perturbation theory, where we integrate open string loops to all orders.
In fact, the parameter that controls this perturbative expansion 
is $R_s/r$, so we will consider very large $r$: we are expanding around
flat spacetime. Furthermore, for $R_s^2>>\alpha'$, it is possible to neglect
the effects of higher string modes, even in the near-horizon limit.

In principle, since we are including open string loops to all orders, it
would seem that we also need to include closed string loops. However,
closed string loops are suppressed in large $N$: they contribute at order
$g^2\sim (gN)^2\times {1\over N^2}$, and can be safely ignored. Open string loops
contribute at order $gN$ since the worldsheet boundary can be placed at 
any of the $N$ D-branes.

To begin, let us consider the effects of a single boundary.
Polchinski\cite{RRPol} computed the tadpoles of the graviton and RR
4-form potential in the presence of a D-brane to determine that a single
D-brane carried one unit of RR-charge. The same calculation can be used
to deduce that the gravitational field at a distance $r$ from a D3-brane
has strength $\tilde G(X)=R_s^4/2r^4$.
In this calculation, the information about polarizations is lost. This
information may be recovered by a consideration of scattering of
gravitons off of a D-brane.\cite{Klebanov:1996ni,Gubser:1996wt} 
The result is that the polarization manifestly preserves
$SO(6)\times SO(3,1)$,
\beq
g_{\mu\nu}(X)={R_S^4\over 2r^4} t_{\mu\nu}
\equiv{R_S^4\over 2r^4}\left( \delpp_{\mu\nu}-\delpa_{\mu\nu}\right)
\eeq
This is the leading term in an expansion of the supergravity
metric \eq{bgmetric} and we will see that it is consistent from
the point of view of the string path integral as well.

If we are at large $r$, we can think of this as a gravitational
fluctuation, with (on-shell) vertex operator in the (0,0)
picture\footnote{Our normalizations are $\langle
X(z)X(z')\rangle=-{\alpha'\over2}\ln|z-z'|^2$ and
$\langle\psi(z)\psi(z')\rangle=1/(z-z')$. Furthermore, 
$G(X)={\tilde G(X)\over 4\pi}$ (the $4\pi$ is introduced to account for
the differing normalizations of metric and vertex operator.)}
\beqn\label{eq:gravitonvop}
\delta V_g &=& -{2\over\alpha'}t_{\mu\nu}
\left(-G(X)\paX^\mu\bpaX^\nu
+{\alpha'\over 2}(\psi\cdot\pa G(X))\psi^\mu\bpaX^\nu\right.\nonumber\\
& &\left.+{\alpha'\over 2}\paX^\mu(\tilde\psi\cdot\pa G(X))\tpsi^\nu
-\left({\alpha'\over 2}\right)^2 \psi^\lambda\psi^\mu\tpsi^\rho\tpsi^\nu
\pa_\lambda\pa_\rho G(X)\right)
\eeqn
We will find that it is necessary to modify this vertex operator by
contact terms; we will see that this is equivalent to going to $N=1$
superspace.

The RR vertex operator in the $(-{1\over2},-{1\over2})$-picture takes
the form\footnote{$C$ is charge conjugation, $\Gamma_{-1}$ is
10-dimensional chirality, and $\Gamma_{\perp(\parallel)}$ is chirality
in the 6(4)-dimensional subspace.}
\beql{RRvop}
\delta V_{RR}
=\tilde S^T C\Gamma^{\mu_1\ldots\mu_5}\left( {1+\Gamma_{-1}\over 2}\right) S
H_{\mu_1\ldots\mu_5}e^{-\phi/2}e^{-\bar\phi/2}
\eeq
where
\beq
\Gamma^{\mu_1\ldots\mu_5}H_{\mu_1\ldots\mu_5}
=\Gamma_\parallel\Gamma^\alpha\pa_\alpha G(X)
\eeq
The normalization of $V_{RR}$ may be determined by factorization. 

The first step is to check under which conditions the operator
\eq{gravitonvop} is well-defined; instead of defining it to be
normal-ordered, we require that the singularities are absent. There are
both quadratic divergences (contact terms)\cite{MGNS} and logarithmic divergences
to consider. We will confine our attention to the $\paX\bpaX$ part of
the graviton vertex operator. The other terms follow from
worldsheet supersymmetry.

The prefactor $G(X)$ will be dealt with by expansion around a fixed 
configuration:
$G(X)=\langle G(X)\rangle+X^\mu\langle \pa_\mu G(X)\rangle+\ldots$.
Point-splitting $\delta V_g$ (eq. \eq{gravitonvop}), we find
\beqn\label{eq:selfcontract}
\int d^2z\ \delta V_g=\int d^2z\ :\delta V_g: + 
\tr t\ 2\pi\delta^{(2)}(\epsilon)\int d^2z :G(X):\\
-{1\over 2}\ln |\epsilon|^2\ \int d^2z :\Box G(X) t_{\mu\nu}\paX^\mu\bpaX^\nu:
+\ldots\nonumber\eeqn
In the ellipsis are the fermion terms. There are higher order terms in $\alpha'$,
but they are all proportional to (derivatives of) $\Box G$.
Now, because $G(X)$ is a harmonic function, $\Box G(X)$ vanishes, at least away
from the branes. Thus $\delta V_g$ is defined up to a contact term. This
contact term may be removed by modifying $\delta V_g$ by terms which vanish
on the (tree-level) equations of motion. In fact, these modifications are
equivalent to introducing $N=1$ superspace
\beq
{\bf X}=X+i\theta\psi+i\bar\theta\tilde\psi+\theta\bar\theta F
\eeq 
whereupon the vertex operator is written:
\beql{newgop}
\delta V_g=t_{\mu\nu}\int d^2z\int d^2\theta\ G({\bf X})D{\bf X}^\mu\bar D{\bf X}^\nu.
\eeq
The new terms in the vertex operator 
which are relevant for the contact terms are
\beq
Gt_{\mu\nu}(\psi^\mu\bar\pa \psi^\nu+\tilde\psi^\nu\pa\tilde\psi^\mu+F^\mu F^\nu)
\eeq
The fermionic terms in the vertex operator are just the appropriate covariantizations
\beq
\psi^\mu(G\bar\pa \delta^\nu_\lambda +\bar\pa X^\nu \pa_\lambda G)\psi^\lambda,
\eeq
etc. These give additional contractions which cancel the contact term.

Thus we have shown that the vertex $\delta V_g$ is well-defined. The RR
vertex operator requires no such treatment, as there are no contact
terms;\footnote{This assumes the spin field $S^\alpha$ is normal-ordered.}
 it is regular, as long as $\Box G=0$. With these operators in
hand, we are prepared to consider string worldsheets with multiple
insertions. We will find that there are additional logarithmic
short-distance singularities, which can be canceled by the
Fischler-Susskind mechanism, namely by modifying the function $G$. This
calculation appears in Section 3.

Let us consider the scattering of string states off of the D-branes
at low momentum transfer. The momentum transfer is the Fourier transform of
the coordinate $r$. We want to show that the contributions to this 
scattering are dominated by the exchange of massless states. The boundary
on the D-brane may then be replaced in this factorization limit, by an
insertion of  a massless background vertex.

Indeed, this is what we would expect from looking at a Born-Oppenheimer
approximation to the scattering. In essence, if each boundary has this
property, then the relevant limit for calculating amplitudes involves
long thin tubes attached to the brane, so each boundary is effectively
shrunk to zero size, and can be replaced by a local insertion of a
massless vertex operator, which depends on $r$. This
dependence comes about as each long thin tube extends from where the
interaction takes place to the D-brane, and hence involves the Green's
function $\sim {1\over r^4}$ for the massless state in consideration.

The fact that the amplitude factorizes into a long thin tube for each
boundary  is a non-trivial statement, as one might imagine that there
might be hard momentum flowing from one hole to another, thus making the
boundaries big. We will argue that these effects cancel each other by
supersymmetry. Indeed, in a low momentum scattering, the amplitude of
external states factorizes onto a one point function of a vertex
operator that is almost on-shell and massless, on a Riemann surface with
$ m>1$ holes. As we have more than one hole in the surface, the diagram
without insertion is a self energy diagram of the brane configuration,
and by supersymmetry vanishes; addition of the massless state insertion
to this diagram is zero on-shell, as it probes the self-energy of the
brane externally. The argument breaks down  for $m=1$, as this probes
the 'tree' level contribution to the mass of the brane, and this is the
D-brane tension. Thus, the relevant limit of moduli space that
contributes to the amplitude involves long thin tubes for all the
insertions, and hence we get only effects from massless vertex operators
in the full calculation.

Summing over this restricted  moduli space of zero size holes 
integrates these vertex operator insertions over a Riemann surface with
fixed complex structure. We sum
over all inequivalent surfaces, and since the boundaries are all
identical, we should divide out by the permutation symmetry to
avoid overcounting. 
The sum over boundaries generates the series
\begin{equation}
1+\sum_{m=1}^\infty \frac 1{m!}\left(\int_\Sigma  V_B \right)^m = \exp{\int_\Sigma d^2 z V_B}
\end{equation}
with $V_B$ a massless string state generated by the boundary.
That is, the sum over boundaries generates an expression which has a sigma model
interpretation where we modify the action by 
\beql{genexp}
S\to S+\int_\Sigma d^2 z V_B(r)
\eeq
where $r$ is the impact parameter quantum field on the worldsheet. 
The sum over spin structures implies that 
$V_B$ is given by the combined graviton and self dual RR tadpole
which we extracted in eqs. \eq{newgop},\eq{RRvop} above.

From the point of view of 
string perturbation theory around flat spacetime, this corresponds to
a partial resummation. As far as the metric is concerned,
the exponentiation is standard. For the RR background, as discussed,
the combinatorics are right for exponentiation, but the superconformal
ghost factors make this problematic. A $\sigma$-model for these backgrounds
may be written, formally, in several ways. The basic issue is that 
we wish to write an expression with zero ghost charge. In an S-matrix
element, it would be convenient to take the external states in a fixed
picture, and the insertion of eq. \eq{genexp} implies that each term
in $V_B$ should be in the same picture.

There may exist a field redefinition which mixes ghosts and matter 
fields similar to Ref. \cite{Berkovits:1994wr,BVW} that would sidestep this problem
but instead we propose to formally write the $\sigma$-model as
\beq
S_{RR}\sim \int P_{1/2}\bar P_{1/2} V_{(-1/2,-1/2)}
\eeq
where $P_{1/2}$ ($\bar P_{1/2}$) increases the L(R)-ghost charge by $1/2$ and 
satisfies $P_{1/2}^2=P_{+1}$ ($\bar P_{1/2}^2=\bar P_{+1}$). Such
operators are clearly not well-defined. However, this expression has two
important properties: {\it it generates the correct combinatorics, and is
unambiguous for any non-zero S-matrix element}. The basic point is that
in an expansion of this exponential, the RR background vertex only 
contributes in pairs (on any topology). Using this $\sigma$-model,
one can systematically compute perturbative corrections. For a given
on-shell S-matrix element on a given topology,
one introduces the correct background ghost charge explicitly, then takes
each vertex operator, and each background vertex in the $(0,0)$ picture,
defined as above. Further issues involving the properties of $P_{1/2}$
may be found in Sections 5,6.

An alternative procedure would be the following.
Since the one-point function of $V_{RR}$
is zero on any topology, one could attempt to write non-local expressions
of the form
\beq
S_{RR}\sim \int V_{(+1/2,+1/2)}\int V_{(-1/2,-1/2)}+\ldots
\eeq
but the combinatorics of such an action are not correct: one needs to
correct this at order $V^4$ and so on. An action with an infinite number
of non-local terms is certainly not a terribly convenient representation.

\section{Fischler-Susskind and the $\sigma$-model}

Having discussed the general form of string perturbation theory in this
background, we now give an explicit calculation of the modifications to
the background which follow from conformal invariance of the string path
integral. The calculation is an application of the Fischler-Susskind mechanism
for a pair of colliding worldsheet holes. Thus we consider two background
vertex operators, defined above, and bring them close together.

\smallskip
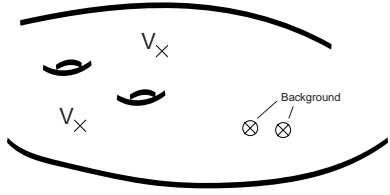
\centerline{\littlefig{opprodFS.EPSF}{2}}

\noindent The calculation of the short-distance singularity is somewhat involved
but straightforward, and
we present here only the result. The quadratic divergences (contact terms)
cancel exactly, given the modified form of the vertices discussed above. 
For simplicity, we write here the contributions to only the $\paX\bpaX$ part
\beqn
\int d^2z\ \delta V_g(z)\cdot\int d^2z'\ \delta V_g(z')=&&\nonumber\\
-4\pi\ln |\epsilon|^2\int d^2z
\Bigg(\paX^\mu\pa_\mu G\ \bpaX^\nu (t^2)_\nu^{\ \lambda}\pa_\lambda G\\
+\bpaX^\mu\pa_\mu G\ \paX^\nu (t^2)_\nu^{\ \lambda}\pa_\lambda G
\nonumber\\
 -{1\over2}(\tr t^2) \paX^\mu\pa_\mu G\ \bpaX^\nu\pa_\nu G
-\paX^\mu t_\mu^{\ \rho}\pa_\rho G\ \bpaX^\nu t_\nu^{\ \lambda}\pa_\lambda G\nonumber\\
+(\pa G\cdot \pa G)(t^2)_{\mu\nu}\paX^\mu\bpaX^\nu
+G\Box G\ t_{\mu\nu}\paX^\mu\bpaX^\nu\Bigg)\nonumber
\eeqn
Since $G(X)$ depends only on $r$, using the form of $t_{\mu\nu}$ discussed
earlier this reduces to
\beqn
\int d^2z\ \delta V_g(z)\cdot\int d^2z'\ \delta V_g(z')=\nonumber\\
4\pi\ln |\epsilon|^2\int d^2z 
\left(4\pa_\mu G\pa_\nu G-(\pa G\cdot \pa G)\eta_{\mu\nu}\right)
\paX^\mu\bpaX^\nu+\ldots
\eeqn
which is in the form of a graviton vertex operator.

There is also a logarithmically divergent contribution of the same form
from two background RR vertices. After some Dirac algebra, we find
\beqn
\int d^2z\ \delta V_{RR}(z)\cdot\int d^2z'\ \delta V_{RR}(z')=\nonumber\\
4\pi\ln|\epsilon|^2\int d^2z 
\left(-4\pa_\mu G\pa_\nu G+2(\pa G\cdot \pa G)t_{\mu\nu}\right)
\paX^\mu\bpaX^\nu
\eeqn
To obtain this result, the normalization of the RR vertex operator is
required, as given in \eq{RRvop}, and discussed in Section 5. 

The total logarithmic divergence in the graviton channel is then of the form
\beql{totdiv}
4\pi\ln |\epsilon|^2\int d^2z 
(\pa G\cdot \pa G)\left( \delpp_{\mu\nu}-3\delpa_{\mu\nu}\right)
\paX^\mu\bpaX^\nu+\ldots
\eeq
Note that the terms in $\pa r\bar\pa r$ have cancelled precisely between
NS and R contributions.\footnote{Even if this were not true, a field 
redefinition can remove such terms.} In this sense, we have started in the
``right" coordinate system. 

The divergence \eq{totdiv} may be removed by modifying the
graviton background through the addition of a term
\beq
-2\pi \left( {2\over \alpha'}\right) G^2(X)
\left( \delpp_{\mu\nu}-3\delpa_{\mu\nu}\right)
\paX^\mu\bpaX^\nu+\ldots
\eeq
where the ellipsis contains fermionic terms, determined by
supersymmetry. These terms have a logarithmically divergent
self-contraction which exactly cancels that of eq. \eq{totdiv}. To see
this, one can use eq. \eq{selfcontract} with $G$ replaced by $-2\pi
G^2$: since $G^2$ is not harmonic, there is a non-zero contribution from
$\Box G^2$. 

At this order then, the graviton background has the form
\beqn
{1\over 2\pi\alpha'}\left\{ \left(\tilde G-{1\over2}\tilde G^2\right)\delpp_{\mu\nu}
+\left(-\tilde G+{3\over2}\tilde G^2\right)\delpa_{\mu\nu}\right\}
\paX^\mu\bpaX^\nu+\ldots\\
\simeq{1\over 2\pi\alpha'}
\left\{ (f^{1/2}-1)\delpp_{\mu\nu}+(f^{-1/2}-1)\delpa_{\mu\nu}\right\}
\paX^\mu\bpaX^\nu+\ldots
\eeqn
where $f=1+2\tilde G(r)$ is the harmonic function appearing in the
supergravity metric. Thus we see that at lowest order, the
Fischler-Susskind mechanism does build up properly the background
geometry of the 3-branes. Similar calculations may be performed without
difficulty to demonstrate the appropriate modification of the RR
background. We believe that this works to all orders in ${R_s\over r}$;
comments in this regard may be found in a later section of this paper.
Certainly, this is not in disagreement with spacetime arguments that the 
near-horizon geometry is exact.\cite{MGTB,Kallosh:1998qs} 

We present the following worldsheet argument that this is exact to all
orders in $R_s/r$. Since $\alpha'/R_s^2$ is small, it is useful to
consider an expansion of the $\sigma$-model in powers of curvature. In
order to simplify this expansion, one goes to normal coordinates around
a point. In
terms of this expansion in curvature, the Ramond background is a first
order perturbation, and the curvature correction to the metric is second
order. In contrast, for the coordinates we were using before, which are
not normal coordinates, the first correction comes from the Christoffel
symbols, which are set to zero locally on the normal coordinate system.

This can be done for the full geometry, but for simplicity, we specialize
now to the near horizon region.
For $AdS_5\times S^5$, expanding around zero in the normal coordinates we have
\begin{equation}
ds^2 \sim \eta_{\mu\nu}dx^\mu dx^\nu + A x_{AdS}^2 dx^2_{AdS}
-A x^2_{S_5} dx_{S_5}^2+\dots
\end{equation}
where $A$ is determined by the curvature of the space.
Also 
\begin{equation}
V_{RR} =P_{1/2}\bar P_{1/2} S^\alpha \tilde S^\beta h_{\alpha\beta}
\end{equation}
with $h_{\alpha\beta} \sim [C(\Gamma_{AdS} + \Gamma_{S_5})]_{\alpha\beta}$.
The $\Gamma$ matrices used are products of the five gamma matrices tangent
to $AdS_5$ and $S_5$ respectively.

Again we consider the collision of two background vertices.
The algebra is straightforward, 
and we proceed by cancelling the logarithmic corrections from the square of the 
Ramond background, and the curvature self-contraction. This produces the 
beta function for the graviton, and it reads
\begin{equation}
(R_{\mu\nu} + (H^2)_{\mu\nu})\log(|\epsilon|^2)=0
\end{equation}
We have used the $P_{1/2}$ operators for the Ramond vertex. As we
have two Ramond vertex operators and to leading order they are on-shell
(they are constant), the factors of $P_{1/2}$ give us a picture-changing
operator, and the
result is unambiguous. We thus reproduce the supergravity equations of
motion for the graviton. This computation generalizes to any
weakly coupled string theory at small curvature in the presence of RR
backgrounds and gives the supergravity equations of motion.

This computation is one-loop in the $\sigma$-model and thus is correct
to leading order in 
$\alpha'$. Again according to \cite{MGTB,Kallosh:1998qs}, this result is
exact for the near-horizon geometry to all orders in $\alpha'$.


\section{Spectrum: Mixing of RR and NSNS states}

We have suggested in the above discussions that the $\sigma$-model may
be consistently defined, and it is exactly conformal. The next step
would be to discuss the spectrum of excitations around the background.
Since the $\sigma$-model is known and exactly conformal, in principle
one could write equations for vertex operators which make them of
dimension $(1,1)$. The best we can do at present is to compute vertex
operators for these excitations perturbatively in the normal coordinate
expansion, starting with flat spacetime operators.

Note that in the approximation in which we are working, the spacing of
states is as shown in the accompanying figure. The gap between levels is of order 
$\sqrt{\alpha'}$ while the spacing between $S^5$ harmonics is of order
$1/R_s$. 
\smallskip
\centerline{\littlefig{spectrumlevels.EPSF}{3}}

We expect this spectrum to be modified by the background. The
leading effect is to mix states at the same level. Because of the RR
background, the states which in flat space are RR and NSNS mix with one
another, which means that there are no longer separately conserved number
operators in the R- and NS-sectors. At most, we can expect that a linear
combination is still a good quantum number, and thus organizes the spectrum,
although it seems unlikely that even this is true at finite $N$. 

Let us consider the one-loop partition function. In flat spacetime, bosonic
symmetries which act on states come in the NS sector, and thus spacetime
symmetries may be thought of as acting on, for example, the NSNS part of
the partition function and the RR part {\it separately}. Once we turn on
the background, this can no longer be true: there will be spacetime 
symmetries in the RR sector and thus at best we can classify states as
bosonic or fermionic (which cancel by supersymmetry).

In this section, we simply demonstrate this mixing at lowest order in
$\alpha'$. The mixing is of order one. In principle, one can
systematically compute higher order corrections. The basic idea is again
an application of the Fischler-Susskind mechanism. We consider some
arbitrary $(1,1)$ operator ${\cal O}e^{ip\cdot X}$ inserted on the
worldsheet. We compute the effects of the background perturbatively by
considering the effects of background vertex operators.

\smallskip
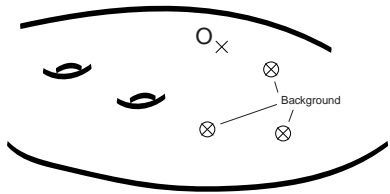
\centerline{\littlefig{spectrumFS.EPSF}{2}}

The leading effect will come from 
a single RR insertion
\beq
\delta V_{RR}=\int d^2z\ P_{1/2}\bar P_{1/2}
(CH)_{\alpha\beta}\tilde S^\alpha S^\beta e^{-\phi/2}e^{-\bar\phi/2}
\eeq
The operator product of $\delta V_{RR}$ with the vertex ${\cal O}$ will be of the form
\beq
\int d^2z\ {1\over |z-z'|^2} (CH)_{\alpha\beta}
\left[ \bar Q^\alpha Q^\beta {\cal O}e^{ip\cdot X}\right]
\eeq
This is logarithmically divergent, and converts NS to R and vice versa.
If need be this can be brought back to, say, the $(0,0)$ or
$(-{1\over2},-{1\over2})$-picture by applying the picture-changing
operator.\footnote{Note that at lowest order, we can assume that the
picture changing operator is unchanged from its flat-space value. Higher
order corrections must take this consistently into account.}

Thus the leading order effect is to mix NS with R:
\beq
\delta\pmatrix{{\cal O}_{NS}\cr {\cal O}_R}= 
\pmatrix{0&M\cr \tilde M& 0}\pmatrix{{\cal O}_{NS}\cr {\cal O}_R}
\eeq
where $M$ and $\tilde M$ are suitable (logarithmically divergent)
mixings that may be computed from the operator products mentioned above
(we consider an explicit example below.) In order to eliminate the
divergence, we should modify the vertex operators, in two steps. First,
we diagonalize
\beq
{\cal O}_D\sim \alpha {\cal O}_{NS}+\beta {\cal O}_R
\eeq
and then modify the operator ${\cal O}_D$ to eliminate the divergence. 
 In the weak curvature limit, it
is sufficient to consider
\beq
:{\cal O}_D\ e^{ik\cdot X}:\to :{\cal O}_De^{ik\cdot X}: e^{i\delta k\cdot X}
\eeq
Going beyond the weak curvature approximation is a difficult problem,
equivalent to solving for the explicit form of the vertex operators in
this background, and we do not attempt such an analysis here. The
modification of the momenta accounts for the change in the mass of states in
the background and is the only source of a logarithmic divergence within
the  vertex operator.

In principle, this analysis may be carried out for any state in the
string spectrum, although it is technically challenging. We confine the
discussion to a demonstration of the mixing effect for components of the
massless gravity multiplet. For these states, the analysis reduces to
that of Ref. \cite{Kim:1985ez} for supergravity in $AdS_5\times S_5$.

Begin then with massless states:
\beqn
{\cal O}_{NS}&=&\int d^2z\ h^{(NS)}_{\mu\nu}(-\paX^\mu+ik\cdot\psi\psi^\mu)
(-\bpaX^\nu+ik\cdot\tilde\psi\tilde\psi^\nu)
e^{ik\cdot X}\\
{\cal O}_R&=&\int d^2z\ h^{(R)}_{\alpha\beta}\tilde S^\alpha S^\beta 
e^{-\phi/2}e^{-\bar\phi/2}e^{ik\cdot X}
\eeqn
We note that
\beq
V_B\cdot {\cal O}_{NS}
=-\int d^2z'\ e^{-\phi/2}e^{-\bar\phi/2}e^{ik\cdot X}S^\alpha\tilde S^\beta\cdot
2\pi\ln\epsilon\ h^{(NS)}_{\mu\nu}k_\lambda k_\rho 
(C\Gamma^{\lambda\nu}H\Gamma^{\rho\mu})_{\alpha\beta}\nonumber
\eeq
and
\beq
V_B\cdot {\cal O}_{R}
=\int d^2z\ 2\pi\ln\epsilon\ h^{(R)}_{\alpha\beta} (C\Gamma^\mu H\Gamma^\nu)^{\alpha\beta}
\psi^\mu\tilde\psi^\nu e^{-\phi}e^{-\bar\phi}
\eeq

These results are equivalent to a mixing matrix of the form:
\beq
-2\pi\ln\epsilon\ \pmatrix{0&h^{(NS)}_{\mu\nu}k_\lambda k_\rho
(C\Gamma^{\lambda\nu}H\Gamma^{\rho\mu})_{\alpha\beta}\cr
-h^{(R)}_{\alpha\beta} (C\Gamma^\mu H\Gamma^\nu)^{\alpha\beta} &0}
\eeq
and agrees with the supergravity analysis of \cite{Kim:1985ez}. This 
mixing suggests that the spacetime symmetries will not respect the
splitting between R and NS.

\section{Spacetime Algebra}

Let us  consider the realization of spacetime symmetries
on the worldsheet and their algebra. In this discussion, there are some
subtleties because of limited knowledge of $P_{1/2}$. However, modulo
these subtleties, we are able to reproduce the correct spacetime supersymmetry
in the curvature expansion around the near-horizon geometry. A fuller discussion
will appear in a subsequent publication.

We know that in the presence of D-branes, half of the supersymmetry is
broken, but is recovered in the near-horizon region as a superconformal
symmetry. Our first task is to recover this fact in string perturbation
theory.

It is natural to define supercharges
\beqn
q^\alpha&=&\oint dz\ S^\alpha e^{-\phi/2}\\
\bar q^\alpha&=&\oint d\bar z\ \tilde S^\alpha e^{-\bar\phi/2}
\eeqn 
which certainly generate symmetries in flat spacetime. 
In a RR background, at lowest order in $\alpha'$, there is only one
linear combination 
\beql{mixedsuperch}
Q^\alpha\sim P_{1/2}q^\alpha+\Gamma_\parallel\bar P_{1/2}\bar q^\alpha
\eeq
retained. To see this, we simply need to require that the supersymmetry
annihilates the background
\beql{bgsusy}
Q^\alpha\cdot V_{B}=0
\eeq
This calculation in fact is another way to fix the normalization of the
RR vertex operator. At lowest order, our lack of understanding of
$P_{1/2}$ is irrelevant. At higher orders, we need to know $P_{1/2}$, 
and furthermore we
expect that eq. \eq{mixedsuperch} is further modified by functions of
$r$. Once this happens, there is no possible splitting between
holomorphic and anti-holomorphic fields.
This effect can be computed from worldsheet considerations (given
$P_{1/2}$) by a further application of Fischler-Susskind. From the
spacetime point of view, they are determined by the requirement that $Q$
be associated to a Killing spinor.

We note the very important point here that the holomorphic and anti-holomorphic
charges mix with one another in the RR background. There is nothing
really exotic about this: given a conserved current $J$, we can write
$\bar\pa J+\pa \bar J=0$, and the spacetime charge
$T=\oint dz\ J+\oint d\bar z\ \bar J$ is well-defined, being independent of
contour. Only in special cases are $J$ and $\bar J$ separately conserved.

In the near-horizon limit, we expect that exactly this happens, and the
other linear combination of spacetime supercharges $S^\alpha$,  comes
back. Indeed, in normal coordinates for the near-horizon case, we begin
with 32 supercharges, and eq. \eq{bgsusy} does not eliminate any. Instead,
\eq{bgsusy} defines the modifications to the supercharges in the presence
of the background. We may write the Killing spinor equation in the form
\beql{killing}
\nabla Q+Q\cdot V_B=0
\eeq
which is the more general form of \eq{bgsusy}. Consider solving this 
equation order by order in the curvature expansion:
\beqn
Q^\alpha&=&Q^\alpha_{(0)}+Q^\alpha_{(1)}+\ldots\\
\bar Q^\alpha&=&\bar Q^\alpha_{(0)}+\bar Q^\alpha_{(1)}+\ldots
\eeqn
The lowest order terms are the flat space expressions 
$Q_{(0)}=q$, $\bar Q_{(0)}=\bar q$. At next order, the RR background in
$V_B$ gives a non-zero contribution to the second term in eq. \eq{killing}
which is proportional to $\bar P_{1/2}\bar q^\alpha$. Since in the normal
coordinate expansion, the Christoffel symbols vanish, \eq{killing} then
implies that 
\beq
Q_{(1)}\sim \oint d\bar z \bar P_{1/2} X\cdot \Gamma\tilde S
\eeq
with a similar expression for $\bar Q_{(1)}$. Thus we see that indeed the
supercharge has both holomorphic and antiholomorphic contributions. This
is precisely what we need to reproduce the spacetime algebra. With the
expressions that we have given, one may check the following commutators are
induced by the worldsheet representation:
\beqn
\{ Q^\alpha, \bar Q^\beta\} &\sim & \oint (\Gamma_{\mu\nu})^{\alpha\beta}
(X^\mu\cdot \paX^\nu-X^\mu\cdot\bpaX^\nu)+\ldots\\
\{ Q^\alpha, Q^\beta\} &\sim & \oint (\Gamma_\mu)^{\alpha\beta}\paX^\mu+\ldots
\eeqn

In general, in order to compute such effects precisely, we need to
understand in more detail the operator $P_{1/2}$. We believe that these
obstacles can be overcome.

\section{Conclusions and discussion}

In this paper, we have analyzed the background produced by D3-branes. We
have discussed how a $\sigma$-model description can be used. The background
geometry is recovered by summing over boundaries of worldsheets, and 
systematically applying the Fischler-Susskind mechanism. Our results 
suggest to us that an exact string conformal field theory description
may exist. In order to `derive' such a description from perturbation
theory requires further understanding of the square root of the
picture changing operator. Our results indicate that this is not such
an unnatural object. With this operator we can construct a perturbation
expansion which is independent of the zero mode of the bosonized superconformal
ghost system. This indicates that picture-changing is still a property of
these $\sigma$-models, but the detailed form is modified by the background.

It is plausible that the approach of \cite{BVW} is related to these remarks.
The case of $AdS_3$ discussed there is considerably different however. In
particular, there are half as many supersymmetries as for $AdS_5$, and
consequently more freedom to choose pictures, and as well the target space
of the $\sigma$-model is a group manifold. It seems difficult
to imagine that a free-field realization exists in that the mixing of
holomorphic and anti-holomorphic currents would seem to invalidate the
idea of a spectrum generating algebra.

\medskip

\noindent {\bf Acknowledgments:} We wish to thank F. Larsen for
collaboration at an early stage of this work. Research supported in part
by the United States Department of Energy grant DE-FG02-91ER40677 and an
Outstanding Junior Investigator Award.

\providecommand{\href}[2]{#2}\begingroup\raggedright\endgroup
\end{document}